# Unusual Resistive Transitions in the Nodal-Line Semimetallic Superconductor NaAlSi

Daigorou Hirai[1], Toshiya Ikenobe[1], Takahiro Yamada[2], Hisanori Yamane[2], and Zenji Hiroi[1]

[1]*Institute for Solid State Physics, University of Tokyo, Kashiwanoha 5-1-5, Kashiwa, Chiba 277-8581, Japan*

[2]*Institute of Multidisciplinary Research for Advanced Materials, Tohoku University, Katahira 2-1-1, Aoba-ku, Sendai 980-8577, Japan*

NaAlSi is a quasi-two-dimensional semimetal with superconductivity below $T_c$ = 6.8 K and a band structure characterized by nodal lines near the Fermi level and potential topological surface states. Electrical resistivity measurements on its superconducting transitions in magnetic fields were made using plate-like single crystals. In the magnetic field–temperature phase diagram, we observed a substantial reduction in resistivity in a pre-transitional zone above the bulk superconducting regime only when the magnetic fields were perpendicular to the plane, rather than parallel to it. Significant sample (thickness) dependence, reentrant behavior, and sensitivity to electrode configurations all indicate that a portion of the crystal has an upper critical field greater than the bulk superconductivity in the pre-transitional region. This fractional superconductivity may occur on the side surface of the crystal.

## 1. Introduction

Topological insulators are states of matter that cannot be connected adiabatically to conventional insulators. They have garnered considerable attention due to their remarkable features, which includes a robust metallic surface state that is protected by the topology having a crossing point of linearly dispersive bands.[1,2] On the other hand, topological semimetals, such as Dirac and Weyl semimetals, have comparable bulk crossing points and may also host topologically protected surface states. Additionally, nodal-line semimetals having extended band crossing points along a specific line or ring were theoretically focused[3,4] and realized in actual materials such as ZrSiS,[5,6] PbTaSe$_2$,[7-9] and CaAgP.[10,11] The "drumhead" form of the topological surface state of such a nodal-line semimetal results in a high density of states (DOS). Thus, the surface state is likely to exhibit exciting properties such as high-temperature superconductivity, flat-band ferromagnetism, and other electron correlation-related phenomena.[12-15]

PbTaSe$_2$ is a rare superconductor with a $T_c$ of 3.72 K in the family of nodal-line semimetals.[7-9] Topological surface states were observed by angle-resolved photoemission spectroscopy[8] and quasi-particle scattering interference imaging in scanning tunneling spectroscopy (STS).[9] Moreover, the STS detected a zero-bias conductance peak at vortex cores in the superconducting state, which was attributed to the Majorana zero energy mode predicted for topological superconductors.[1]

We will focus on another nodal-line semimetal, NaAlSi in this paper. It crystallizes in a layered structure of the anti-PbFCl type with the centrosymmetric space group $P4/nmm$,[16] as depicted in Fig. 1, in which the Al and Si atoms are covalently bonded to create conducting layers separated by layers of the ionic Na atoms. According to pioneering first-principles electronic state calculations,[17] NaAlSi is a self-doped, quasi-two-dimensional semimetal with nearly free electrons and covalent holes: the electronic states near the Fermi level are characterized by an overlap of the highly dispersive electron-like Al 3$s$ bands and the less dispersive hole-like Si 3$p$ bands.

Resistivity measurements at 300 K on single crystals revealed a moderate two dimensionality with anisotropy of approximately 10.[18]

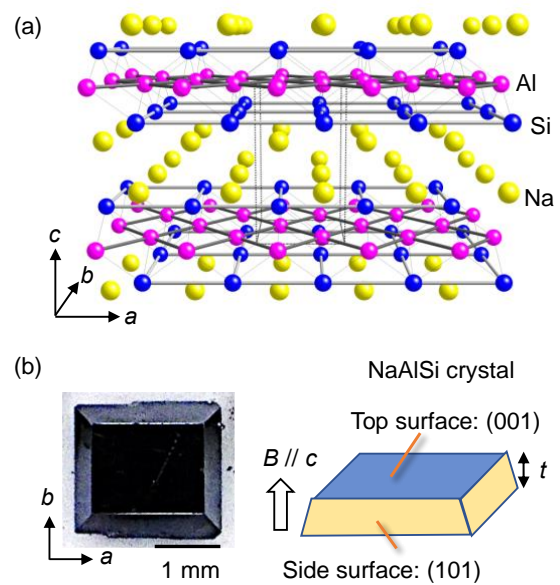

Fig. 1. (Color online) (a) Layered crystal structure of NaAlSi with a tetragonal unit cell (broken lines) of $a$ = 4.1217(1) Å and $c$ = 7.3629(2) Å.[18] (b) A typical crystal with top/bottom surfaces of {0 0 1} and side surfaces of {1 0 1} with a thickness of $t$.

Recent electronic state calculations by the three groups revealed that band overlap results in multiple linear band crossings near the Fermi level, arising in a complex nodal-line structure in the absence of spin–orbit interactions (SOI);[19-21] however, the inclusion of SOI results in tiny gaps of less than 10 meV at all crossing points, which can be ignored under finite thermal perturbation at the experimental conditions.[21] The computations anticipated nontrivial topological surface states:

a drumhead band at around the Γ point on the (0 0 1) surface and a flat band along the Γ–Z line on the (1 0 0) surface.[21)]

Kuroiwa et al. discovered a superconducting transition with an onset temperature of 7 K in NaAlSi in 2007 by measuring the magnetization of polycrystalline samples prepared under high pressure.[22)] Subsequent research indicated that the superconductivity appeared to be of the conventional s-wave type mediated by phonons.[20,23-25)] However, relatively recent bulk measurements on NaAlSi single crystal revealed the possibility of a complex superconducting gap and electron correlation effects in the normal state properties,[18)] suggesting that something exceptional is still waiting to be discovered. It would be fascinating to see if there is an interplay between the superconductivity and topological electronic states in NaAlSi.

In this paper, we describe unique superconducting transitions in resistivity for several single crystals of NaAlSi. We found a decrease in resistivity in the pre-transitional zone above the bulk superconductivity region in the $B$–$T$ phase diagram for $B \parallel c$, but not for $B \parallel a$. The magnitude of the decrease was very sample-dependent; the thicker the crystal, the greater the reduction. It is proposed that a portion of a crystal, which is likely crystal's side surface, has a higher upper critical field than the bulk superconductivity.

## 2. Experimental

As previously reported, single crystals of NaAlSi were grown in a boron nitride crucible using the Na–Ga flux technique.[18)] They contained a trace amount of Ga (1–2% per formula unit) that was incorporated from the flux during crystal growth and acted as a substitute for isovalent Al. Due to the crystal's susceptibility to decomposition when exposed to moisture in the air, handling was carried out in an argon-filled glovebox. Five crystals of the truncated tetragonal pyramid form with {0 0 1} facets at the top and bottom and {1 0 1} facets on the side were picked up; the top and side planes intersect at roughly 60°; the crystal was easily cleaved along the {0 0 1} plane. Typically, the crystal dimensions were $2 \times 2$ mm$^2$ in the plane and $t$ (mm) in thickness. The resistivity measurements were performed on crystals A, B, C, and D with $t$ = 0.7, 0.4, 1.2, and 0.5, respectively. Another thick crystal E with a mass of 11.25 mg was used for heat capacity measurements.

Resistivity ($\rho$) and heat capacity ($C_p$) measurements were performed in a physical property measurement system (PPMS, Quantum Design Inc.). The resistivity was determined using the standard four probe method with indium metal electrodes on the crystal surfaces; typical silver or gold metal pastes were not suitable as electrodes due to their high contact resistance. For the majority of experiments, a pair of current electrodes was positioned on the crystal's top surface and a current $I$ of 1–5 mA was delivered along the $b$ axis (inset, Fig. 2). On the other hand, two pairs of voltage electrodes were constructed on the top and side surfaces for the measurements illustrated in Fig. 7. Except for the measurements illustrated in Fig. 2(a), magnetic fields of 0–5 T were applied along the $c$ axis.

## 3. Results

### 3.1 Resistive transitions under magnetic fields

The superconducting transitions in resistivity obtained in the previous study[18)] for crystal A with $t$ = 0.7 mm are reproduced in Fig. 2. Selected isothermal resistivity curves at $T$ = 1.8–6.8 K under magnetic fields along the $a$ and $c$ axes are presented; the measurements were performed using the identical electrode setup with the voltage electrodes on the top surface and the PPMS rotator. At 1.8 K for $B \parallel a$, the resistivity stays zero at magnetic fields less than an offset field $B_{c2}^{off}$ of 5.8 T and grows to the normal-state resistivity $\rho_n$ at an onset field $B_{c2}^{on}$ of 8.3 T. With increasing temperature, the resistivity curve shifts roughly parallel to the low-field side, which is a characteristic response of the superconducting transition in magnetic fields. In contrast, at 1.8 K for $B \parallel c$, the first increase in resistivity when the magnetic field increases above $B_{c2}^{off}$ = 0.98 T is followed by the second gradual increase above $B_{c2}^{on}$ = 1.5 T to reach $\rho_n$ at $B^*$ = 2.5 T. Upon heating, the resistivity curves move to the low-field side with the second slope diminishing to vanish. The temperature dependences of the three characteristic fields are displayed in Fig. 2(c). $B_{c2}^{off}$ defines the bulk upper critical field $B_{c2}(T)$. The $B_{c2}(0)$ values for $B \parallel a$ and $c$ are estimated to be 7.21(6) and 1.22(2) T, respectively, with the anisotropy of 6.18.[18)] Note that the variations at $B_{c2}^{on}$ and $B^*$ are not smooth but appear to have kinks, indicating that a certain phase transition rather than a crossover is involved. Moreover, the two-step transition seen solely for $B \parallel c$ should not be attributed to conventional surface superconductivity in a two-dimensional superconductor with an upper critical field $B_{c3}$, as $B_{c3}$ should be greater than $B_{c2}$ for $B \parallel a$ and equal to $B_{c2}$ for $B \parallel c$.[26)]

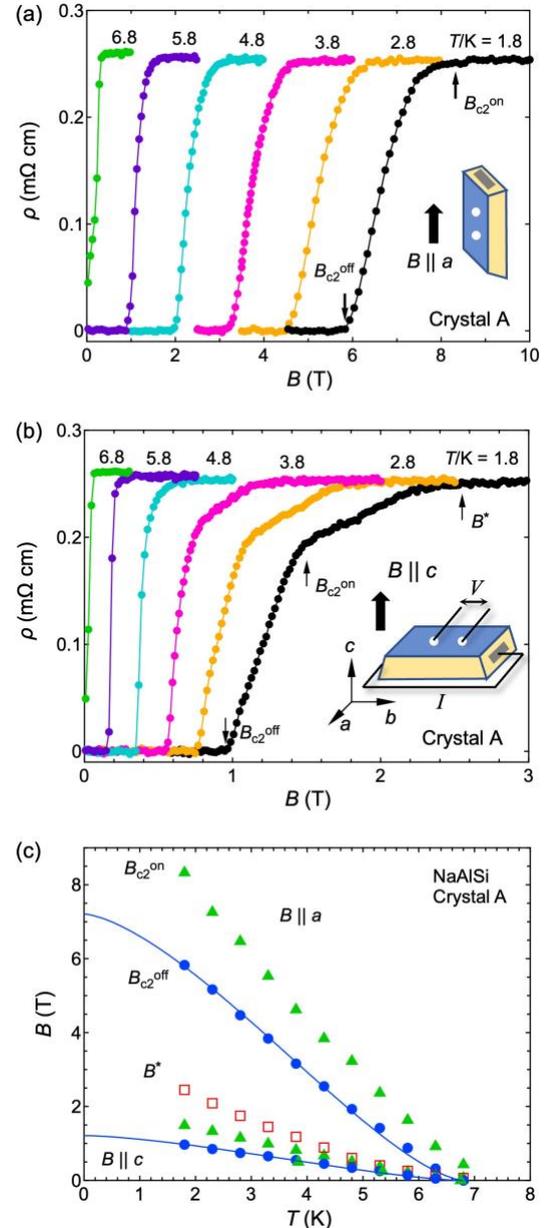

Fig. 2. (Color online) (a, b) Isothermal electrical resistivities of crystal A with $t$ = 0.7 mm.[18)] The measurements were made using a pair of voltage electrodes on the top surface and a current flowing along the $b$ axis under magnetic fields applied along the $a$ axis (a) and the $c$ axis (b), as depicted in the insets. (c) $B$–$T$ phase diagram illustrating the temperature dependences of the characteristic magnetic fields which are commonly described in (a) and (b) for the 1.8 K curves: $B_{c2}^{off}$ (blue circles), $B_{c2}^{on}$ (green triangles), and $B^*$ (red squares). The solid lines are fits to the form $B_{c2}(T) = B_{c2}(0)[1 - (T/T_c)^{3/2}]^{3/2}$, which results in $B_{c2}(0)$ values of 7.21(6) T and 1.22(2) T for $B \parallel a$ and $c$, respectively.

Similar resistivity measurements were made on a thinner crystal B with $t = 0.4$ mm at $B \parallel c$. Figure 3 shows magnetic field and temperature dependences of the resistivity. First, look at the isotherm at 1.8 K in Fig. 3(a). As in crystal A, the zero-resistive state is disrupted above $B_{c2}^{off} = 1.0$ T. As the magnetic field is increased further, the resistivity once approaches to $\rho_n$ at $B_{c2}^{on} = 1.5$ T, falls to form a dip, and then approaches to $\rho_n$ again at $B^* = 2.5$ T. As $T$ increases, this reentrant regime between $B_{c2}^{on}$ and $B^*$ moves to lower fields, changes to a two-step transition at 4.8 K, as in crystal A, and finally vanishes as $T_c(B = 0)$ approaches. The thus-determined characteristic fields are presented in the $B$–$T$ phase diagram of Fig. 4 (circle marks). Notably, despite the significant discrepancies in the behavior of resistive transitions, the three characteristic fields of crystal B coincide with those of crystal A.

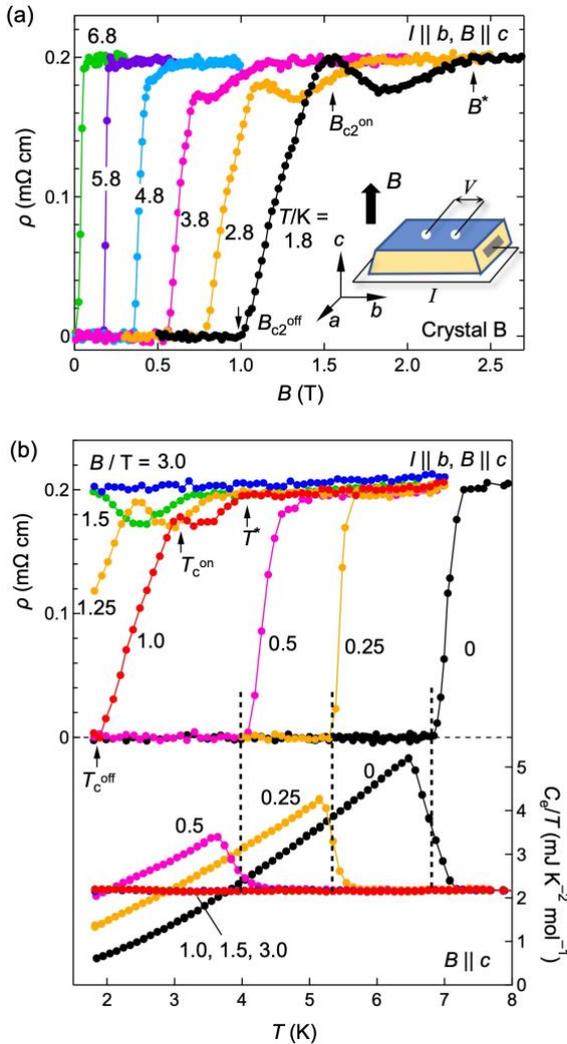

along the $c$ axis from crystal E. $T_c$, $T_c^{on}$, and $T^*$ are commonly shown by the arrows for the 1.0 T curve.

This pre-transitional behavior in crystal B is also evident in the $T$ dependences of resistivity under various magnetic fields in Fig. 3(b). At zero magnetic field, the resistivity abruptly decreases to zero at $T_c(0) = 6.8$ K, which corresponds to the halfway of the leap in the electronic heat capacity recorded on crystal E; this relationship is an empirical signature of a clean superconductor. $T_c$ decreases with increasing magnetic field strength, reaching 5.3 K at 0.25 T and 3.9 K at 0.5 T. Note that the 0.5 T curve has a slight dip at $T^* \sim 5.1$ K before to the abrupt decline. Then, at $B = 1.0$ T, the resistivity begins to decline from $\rho_n$ at $T^* = 4.0$ K and shows a small peak at $T_c^{on} = 3.1$ K, followed by a drop to zero at $T_c(1.0) \sim 1.8$ K; a spike in heat capacity must occur below 1.8 K. Clearly, this complex variation corresponds to the reentrant behavior in the isothermal curves in Fig. 3(a). The 1.5 T curve exhibits a dip between $T^* \sim 3.3$ K and 1.8 K, and the anomaly is eliminated completely at 3.0 T. It is important to observe that the $C_p$ curves at $B = 0.5$–1.5 T do not exhibit any corresponding anomalies at $T^*$. The resulting typical temperatures (triangle marks) are compatible with the characteristic magnetic fields (circle marks) in the phase diagram of Fig. 4; for example, look at the horizontal dashed line at $B = 1.0$ T and the vertical line at $T = 1.8$ K. It is highlighted that bulk superconductivity always exists below $T_c^{off}$ and $B_{c2}^{off}$, but no bulk phase transition happens at $T^*$ and $B^*$.

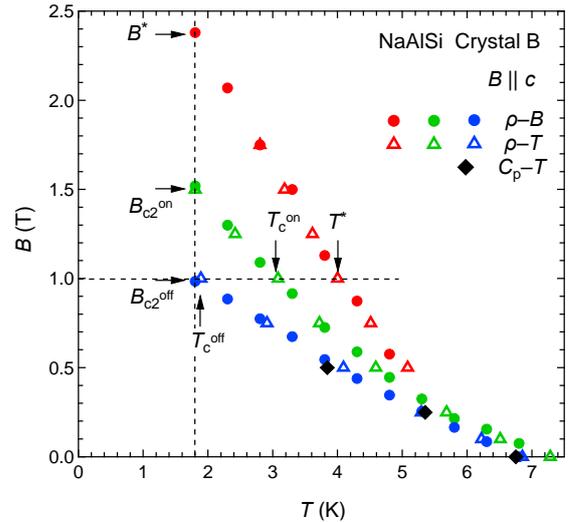

Fig. 4. (Color online) $B$–$T$ phase diagram at $B \parallel c$ obtained using the $\rho$–$B$ data in Fig. 3(a) (circles) and $\rho$–$T$ data in Fig. 3(b) (triangles) for crystal B. The arrows along the vertical broken line indicate $B_{c2}$, $B_{c2}^{on}$, and $B^*$ at $T = 1.8$ K, while those along the horizontal broken line indicates $T_c$, $T_c^{on}$, and $T^*$ at $B = 1.0$ T. The black diamonds also depict the bulk $T_c$ values obtained from the heat capacity data of crystal E in Fig. 3(b).

### 3.2 Sample dependence

We measured resistivity at $B \parallel c$ on thicker crystal C with $t = 1.2$ mm. In comparison to the other two crystals, the $\rho$ of crystal C at 1.8 K increases more gradually above $B_{c2}^{off}$, as seen in Fig. 5. Then comes a rapid climb over $B_{c2}^{on}$ to recover $\rho_n$ at a slightly higher $B^*$ value; the $\rho$ remains finite below $B_{c2}^{off}$ due to an experimental error and should be zero. A comparison of the three crystals reveals a significant sample dependence. Except for the presence of a peak near $B_{c2}^{on}$ for crystal B, the two curves of crystals A and B are nearly identical. It's worth noting that the transition would appear regular in the absence

Fig. 3. (Color online) (a) Isothermal electrical resistivities of crystal B with $t = 0.4$ mm as a function of magnetic field applied along the $c$ axis as measured using top-surface electrodes and an electrical current along the $b$ axis. The arrows indicate $B_{c2}$, $B_{c2}^{on}$, and $B^*$ for the 1.8 K curve. (b) Temperature dependences of resistivity from crystal B and electronic heat capacity divided by temperature $C_e/T$ in magnetic fields

of a dip in thin crystal B. For thicker crystal A, the dip becomes a smooth slope, whereas for more thicker crystal C, the dip becomes a downwardly convex variation with a much smaller magnitude.

These three crystals, as well as the others analyzed in this study, must have almost identical crystalline properties due to their growth in the same sample batch. Indeed, dispersion in their physical parameters, such as $T_c$, $B_{c2}$, normal-state resistivity, and residual resistivity ratio, were negligible. Thus, the sample dependency is most likely due to crystal thickness: the thicker the crystal (in the sequence of crystals B, A, and C), the greater the resistivity reduction in the transitional and pre-transitional regimes.

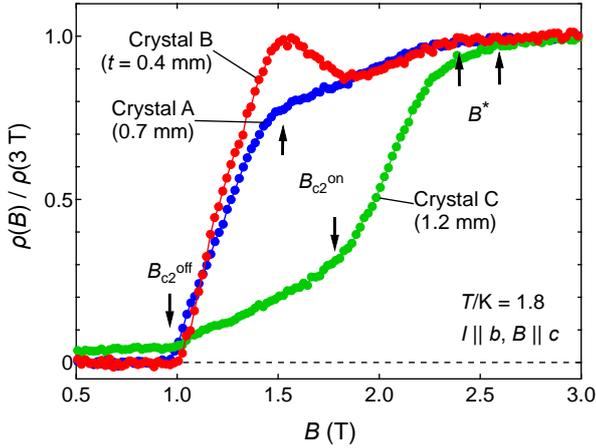

Fig. 5. (Color online) Comparison of the resistivity at 1.8 K for crystals A ($t$/mm = 0.7), B (0.4), and C (1.2) as a function of magnetic field. The resistivities are normalized to their 3 T and 5 K values. The arrows indicate the characteristic magnetic fields of $B_{c2}^{off}$, $B_{c2}^{on}$, and $B^*$. The nonzero resistivity of crystal C below $B_{c2}^{off}$ is an artefact of the experiment.

The $B$–$T$ phase diagram of Fig. 6 summarizes all the characteristic magnetic fields and temperatures from the three crystals analyzed by resistivity measurements, as well as from crystal E via heat capacity measurements. The $B_{c2}^{off}$ values are quite consistent, however there is a slight spread for $B_{c2}^{on}$, which may be owing to the definition's ambiguity. It is now clear that an unexpected pre-transitional zone exists between the $B_{c2}^{on}$ and $B^*$ lines, above the conventional transitional region between $B_{c2}^{off}$ and $B_{c2}^{on}$, leading to bulk superconductivity below $B_{c2}^{off}$.

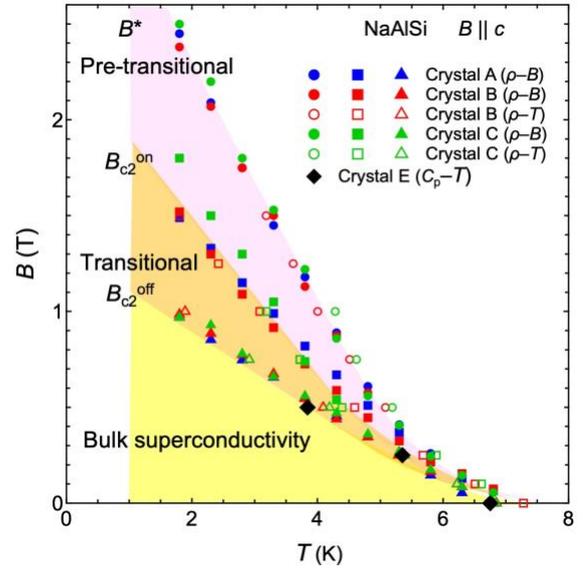

Fig. 6. (Color online) $B$–$T$ phase diagram at $B \parallel c$ for crystal A (blue marks), B (red marks), C (green marks), and E (black marks). $B^*$ (solid circles), $B_{c2}^{on}$ (solid squares), and $B_{c2}^{off}$ (solid triangles) values were determined using $\rho$–$B$ measurements, while $T^*$ (open circles), $T_c^{on}$ (open squares), and $T_c$ (open triangles) values were determined using $\rho$–$T$ measurements. The black solid diamonds indicate the bulk $T_c$ obtained from data on crystal E's heat capacity. Colors denote bulk superconductivity, transitional regime, and pre-transitional regime.

*3.3 Resistivity measurements at the top and side surfaces*

All the resistivity measurements described above were performed using pairs of voltage electrodes attached on the top surface of crystals. To rule out the possibility that the observed atypical transitions were caused by an irregular electrical current flow in a crystal, we repeated the resistivity experiment using two pairs of voltage electrodes on the top and side surfaces of crystal D with $t$ = 0.5 mm. The actual experimental setup is seen in the inset photograph of Fig. 7: a constant current $I$ of 1 mA is sent through the pair of current electrodes covering the facing side surfaces, and voltage drops between the pairs of voltage electrodes on the top ($V_{ts}$) and side ($V_{ss}$) surfaces were detected. The main panel of Fig. 7 depicts the $T$ dependences of "resistivities" that are actually $V_{ts}/I$ and $V_{ss}/I$ after correcting for top and side electrode geometries, respectively.

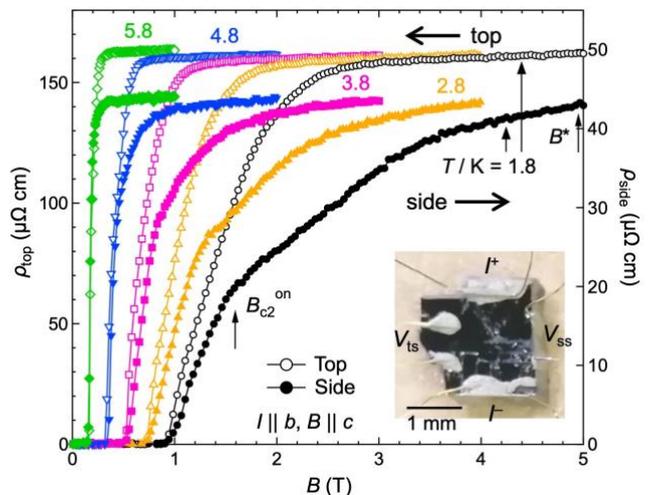

Fig. 7. (Color online) Magnetic field dependences of the resistivity of crystal D with $t$ = 0.5 mm at $B \parallel c$. Two sets of isotherms are shown on the left and right ordinates, using pairs of voltage electrodes on the top ($\rho_{top}$) and side ($\rho_{side}$) surfaces, respectively. The inset photograph

shows the real crystal with a pair of current electrodes ($I^+$–$I^-$) on the side surfaces and two pairs of voltage electrodes ($V_{ts}$ and $V_{ss}$) on the top and side surfaces, respectively. Take note of the enormous magnitude difference in normal-state resistivity.

In comparison to crystal B of comparable thickness (Fig. 3), the two-step transition in the pre-transitional area at $B_{c2}^{on}$–$B^*$ is less noticeable in the top-surface measurements, but is visible in side-surface measurements with much expanded field ranges. Consider a set of curves at 1.8 K that begin to climb identically from zero above nearly equal $B_{c2}^{off}$, then diverge above $B_{c2}^{on}$ and eventually approach $\rho_n$ at $B^* \sim 3$ T and $> 5$ T, respectively, for top- and side-surface measurements. The former $B^*$ value is comparable to those of crystals A, B, and C, but the latter is significantly greater to approach the $B_{c2}$ of 6 T for $B \parallel a$ at 1.8 K [Fig. 2(a)]. Although the $B^*$ values are considerably different, the pre-transitional region was observed in both configurations. This rules out the possibility of a two-step resistive transition being caused by heterogeneous current flow.

It is noted that the magnitudes of $\rho_n$ differ significantly between the two measurements: the $\rho_n$ values at 8 K just above $T_c$ are 172 and 46 $\mu\Omega$ cm for the top and side surfaces, respectively, representing a factor of four discrepancy. Other measurements on three additional crystals revealed comparable variations in $\rho_n$ by several factors. The side surface appears to have a lower resistivity than the top surface, although the reason is unknown.

## 4. Discussion

We observed characteristic resistivity reductions in a temperature (magnetic field) range below $T^*$ ($B^*$) and above the $T_c$ ($B_{c2}$) of the bulk superconducting transition only at $B \parallel c$ in NaAlSi. At this pre-transitional regime, there was a considerable sample (thickness) dependence on the behavior: a smooth variation in resistivity for crystal A, a reentrant change for thinner crystal B, and a larger reduction for thicker crystal C. Similar results were found with two different experimental setups that included voltage electrodes on the crystal's top and side surfaces. In the $B$–$T$ phase diagram, the $B^*$ and $B_{c2}$ lines exhibit comparable $T$ dependences and merge toward zero-field $T_c$ as temperature increases.

It is well established that a low-dimensional superconducting fluctuation results in an increase in conductivity $\sigma$ above $T_c$, as defined by $\sigma \sim (T - T_c)^{-(4-d)/2}$, where $d$ represents dimension.[26] Thus, for two-dimensional superconducting fluctuations, resistivity should decrease linearly with decreasing temperature. Indeed, two-dimensional superconductors such as $MgB_2$[27] and $YBa_2Cu_3O_7$[28] exhibit this property. For example, in a $YBa_2Cu_3O_7$ film, a typical reduction in resistivity was observed below 1.3 times $T_c$. Due to the fact that this effect is essentially a crossover, resistivity begins to fall continuously from a high temperature above $T_c$ and smoothly transitions to a sharp drop at $T_c$. In stark contrast, the observed resistivity variations at $T^*$ ($B^*$) and also at $T_c^{on}$ ($B_{c2}^{on}$) for NaAlSi are not smooth but appear to have kinks, indicating that a certain phase transition with symmetry breaking is involved in this phenomenon.

The relevant phase transition should have occurred non-uniformly or in a small fraction of the crystal, because there was no anomaly at $T^*$ in the heat capacity of Fig. 3(b). Given that the $B^*$ line shows a similar temperature dependence to the $B_{c2}$ line in Fig. 6, it is reasonable to assume that superconducting regions of small volume or thickness with a larger upper critical field than the bulk generate in the matrix, resulting in the reduction in total resistivity. This fractional superconductivity must be the cause of the large sample dependence in the pre-transitional regime. It is emphasized that this fractional superconductivity should not be a consequence of the proximity effect of the bulk superconductivity, as it appears above the bulk superconducting region in the $B$–$T$ phase diagram. Moreover, note that the fractional superconductivity is unconnected to either the surface superconductivity ($B_{c3}$)[26] nor the Fulde–Ferrell–Larkin–Ovchinnikov state[29] for two-dimensional superconductors, as they should occur at $B \parallel ab$, not at $B \parallel c$, in quasi-two-dimensional superconductors.

The merging of the $B^*$ and $B_{c2}$ lines in the phase diagram implies that the zero-field $T_c$ values of the bulk and fractional superconductivities are nearly identical. This indicates that the two superconductivities are based on similar electronic states and the mechanisms are analogous. Thus, it is plausible that NaAlSi itself is responsible for both; provided the high crystalline quality observed in previous single-crystal X-ray diffraction analyses,[18] it is improbable that superconducting impurities with similar $T_c$ values coexisted. Then, a critical question is where in a crystal of NaAlSi the fractional superconductivity occurs. We consider two possibilities: crystal side surfaces and specific crystalline defects inside a crystal. Due to the geometry, when fractional superconductivity occurs in a thin layer that is not exactly perpendicular to the $c$ axis, its $B_{c2}$ can be larger than bulk $B_{c2}(\parallel c)$ and smaller than $B_{c2}(\parallel a)$, such that its contribution appears in a pre-transitional region above the bulk superconducting transition only for $B \parallel c$.

To begin, consider the existence of fractional superconductivity at crystal surfaces. The top surface layer, which is perpendicular to the magnetic field along the $c$ axis, is excluded from the source, whereas the side surface layer, which forms a 30º angle with the $c$ axis, can have an enhanced upper critical field due to the geometry, even if the superconducting properties remain the same. When the bulk superconductivity is suppressed at $B > B_{c2}^{on}$, the side surface layer retains its superconductivity, allowing for increased current flow in the side-surface layer at the expense of bulk current flow. This unidirectional current flow must lead to a lower voltage drop across the top-surface electrodes, as observed by the decreases in apparent resistivity at $B_{c2}^{on}$–$B^*$. The greater reduction observed in thick crystal C in Fig. 5 could be due to the wider side-surface layers contributing more than those of the other crystals. The reason for the absence of a zero resistive state at $B > B_{c2}^{on}$ in the side-surface electrode data could be that the side-surface superconductivity has a broad intrinsic transitional field range due to some reason, or its thickness is extremely thin.

Very interestingly, our recent resistivity measurements on a NaAlSi crystal showed that the resistive transition changes remarkably as a function of magnetic field angle, revealing a distinct two-step transition when the magnetic field is positioned halfway between the $c$ and $a$ axes.[30] The results support the existence of a superconducting surface state on the crystal's side surface to a degree.

If surface superconductivity does exist in NaAlSi, one may consider its relation to the topological surface states.[21] While previous calculations predicted the topological surface states of the (0 0 1) and (1 0 0) surfaces,[21] the topological surface state of the (1 0 1) surface, which serves as the side surface in this study, has not been explored. Although the relationship is intriguing, we currently lack evidence to corroborate it. The surface superconductivity of the topological semimetals $PbTaSe_2$ and $AuSn_4$ has been discussed.[9,31] When bulk superconductivity occurs in these compounds, the surface superconductivity is induced by a proximity effect, and such a fractional superconductivity as observed in NaAlSi was not present. Thus, the possibility of an autonomous superconducting surface state in NaAlSi is intriguing and warrants further investigation.

A similar situation is expected when thin layers surrounding specific crystalline defects exhibit fractional superconductivity. Due to the fact that our crystals contained approximately 1–2% Ga per formula unit, such defects could be caused by Ga impurities. A fractional superconductivity having an enhanced upper critical field could be caused by thin superconducting layers formed around Ga impurities or clusters of defects. Even in this case, a contribution from the interface's topological state may be assumed. In any case, identifying superconductivity that is not bulk is always difficult, and additional careful experiments are required.

Finally, we would like to draw attention to additional mysteries that will be addressed in a future study. The reentrant behavior observed in thin crystal B is unexpected, as it indicates that the contribution of fractional superconductivity below $B^*$ vanishes as the bulk superconducting transition at $B_{c2}^{on}$ approaches. The fractional superconducting region appears to have been absorbed into the matrix, resulting in the loss of the enhanced $B_{c2}$ in its two dimensionality. This implies that there is some "communication" between the surface and bulk superconductivities. On the other hand, the large difference between the top and side electrode measurements in Fig. 7 requires explanation. Additionally, the significant difference in the magnitudes of normal-state resistivity between the top and side electrode measurements is quite strange. On the top (0 0 1) and side (1 0 1) crystal surfaces, it appears as though there are two types of surface states with distinct electronic structures, implicitly implying a relationship with topological surface states. However, we must exercise caution in insisting on this fantastic possibility prior to conducting additional careful experiments. We will conduct additional experiments to elucidate the effect of crystal shape on the intrinsic properties of this fascinating compound, as well as microscopic measurements.

## 5. Conclusions

We observed unusual superconducting transitions in resistivity in the quasi-two-dimensional $sp$ semimetal NaAlSi with topological nodal lines. In the $B$–$T$ phase diagram with $B \parallel c$, there is a pre-transitional zone above the bulk superconducting region where the resistivity decreases in various ways depending on the samples (crystal thicknesses). In the pre-transitional region, it is likely that a fractional superconductivity with enhanced $B_{c2}$ occurs at crystal surfaces or at around certain crystalline defects. NaAlSi provides an intriguing platform for studying a novel aspect of superconductivity.

**Acknowledgments**

The authors wish to express their gratitude to Y. Oikawa, R. Kusaka, and C. Nagahama for assisting with the sample preparation. The authors would like to thank H. Isshiki and Y. Otani for their insightful comments. Additionally, they wish to thank K. Hashimoto and M. Haze for helpful discussions. Financial support for this research was provided by JSPS KAKENHI Grants (JP20H02820 and JP20H05150).

1) M. Z. Hasan and C. L. Kane, Rev. Mod. Phys. **82**, 3045 (2010).
2) X.-L. Qi and S.-C. Zhang, Rev. Mod. Phys. **83**, 1057 (2011).
3) A. A. Burkov, M. D. Hook and L. Balents, Phys. Rev. B **84**, 235126 (2011).
4) S.-Y. Yang, H. Yang, E. Derunova, S. S. P. Parkin, B. Yan and M. N. Ali, Adv. Phys.: X **3**, 1414631 (2018).
5) L. M. Schoop, M. N. Ali, C. Straßer, A. Topp, A. Varykhalov, D. Marchenko, V. Duppel, S. S. P. Parkin, B. V. Lotsch and C. R. Ast, Nat. Commun. **7**, 11696 (2016).
6) S. Pezzini, M. R. van Delft, L. M. Schoop, B. V. Lotsch, A. Carrington, M. I. Katsnelson, N. E. Hussey and S. Wiedmann, Nat. Phys. **14**, 178 (2018).
7) M. N. Ali, Q. D. Gibson, T. Klimczuk and R. J. Cava, Phys. Rev. B **89**, 020505 (2014).
8) G. Bian, T.-R. Chang, R. Sankar, S.-Y. Xu, H. Zheng, T. Neupert, C.-K. Chiu, S.-M. Huang, G. Chang, I. Belopolski, D. S. Sanchez, M. Neupane, N. Alidoust, C. Liu, B. Wang, C.-C. Lee, H.-T. Jeng, C. Zhang, Z. Yuan, S. Jia, A. Bansil, F. Chou, H. Lin and M. Z. Hasan, Nat. Commun. **7**, 10556 (2016).
9) S.-Y. Guan, P.-J. Chen, M.-W. Chu, R. Sankar, F. Chou, H.-T. Jeng, C.-S. Chang and T.-M. Chuang, Sci Adv **2**, e1600894 (2016).
10) A. Yamakage, Y. Yamakawa, Y. Tanaka and Y. Okamoto, J. Phys. Soc. Jpn. **85**, 013708 (2015).
11) Y. Okamoto, T. Inohara, A. Yamakage, Y. Yamakawa and K. Takenaka, J. Phys. Soc. Jpn. **85**, 123701 (2016).
12) N. B. Kopnin, T. T. Heikkilä and G. E. Volovik, Phys. Rev. B **83**, 220503 (2011).
13) R. Nandkishore, Phys. Rev. B **93**, 020506 (2016).
14) Y. Wang and R. M. Nandkishore, Phys. Rev. B **95**, 060506 (2017).
15) Y. Huh, E.-G. Moon and Y. B. Kim, Phys. Rev. B **93**, 035138 (2016).
16) W. Westerhaus and H. U. Schuster, Z. Naturforsch. **34b**, 352 (1979).
17) H. B. Rhee, S. Banerjee, E. R. Ylvisaker and W. E. Pickett, Phys. Rev. B **81**, 245114 (2010).
18) T. Yamada, D. Hirai, H. Yamane and Z. Hiroi, J. Phys. Soc. Jpn. **90**, 034710 (2021).
19) L. Jin, X. Zhang, T. He, W. Meng, X. Dai and G. Liu, J. Mater. Chem. C **7**, 10694 (2019).
20) L. Muechler, Z. Guguchia, J.-C. Orain, J. Nuss, L. M. Schoop, R. Thomale and F. O. von Rohr, APL Materials **7**, 121103 (2019).
21) X. Yi, W. Q. Li, Z. H. Li, P. Zhou, Z. S. Ma and L. Z. Sun, J. Mater. Chem. C **7**, 15375 (2019).
22) S. Kuroiwa, H. Kawashima, H. Kinoshita, H. Okabe and J. Akimitsu, Physica C **466**, 11 (2007).
23) L. Schoop, L. Müchler, J. Schmitt, V. Ksenofontov, S. Medvedev, J. Nuss, F. Casper, M. Jansen, R. J. Cava and C. Felser, Phys. Rev. B **86**, 174522 (2012).
24) H. M. Tütüncü, E. Karaca and G. P. Srivastava, Philos. Mag. **96**, 1006 (2016).
25) H. Y. Wu, Y. H. Chen, P. F. Yin, Z. R. Zhang, X. Y. Han and X. Y. Li, Phil. Mag. Lett. **99**, 29 (2019).
26) M. Tinkham, *Introduction to superconductivity*. (Dover Publications, inc., New York, 1996).
27) A. Sidorenko, L. Tagirov, A. Rossolenko, N. Sidorov, V. Zdravkov, V. Ryazanov, M. Klemm, S. Horn and R. Tidecks, JETP Letters **76**, 17 (2002).
28) M. Putti, M. R. Cimberle, C. Ferdeghini, G. Grassano, D. Marrè, A. S. Siri, A. A. Varlamov and L. Correra, Physica C **314**, 247 (1999).
29) R. Casalbuoni and G. Nardulli, Rev. Mod. Phys. **76**, 263 (2004).
30) T. Ikenobe, D. Hirai, T. Yamada, H. Yamane and Z. Hiroi, J. Phys.: Conf. Ser. (2022).
31) D. Shen, C. N. Kuo, T. W. Yang, I. N. Chen, C. S. Lue and L. M.

Wang, Commun. Mater. **1**, 56 (2020).